\documentclass[aps,prl,twocolumn,showpacs,superscriptaddress,floatfix]{revtex4-1}
\usepackage{graphicx,amsmath,amssymb}
\usepackage[usenames]{color}
\usepackage[dvipsnames]{xcolor}
\usepackage[unicode=true,pdfusetitle,
 bookmarks=false,
 breaklinks=false,pdfborder={0 0 1},backref=false,colorlinks,urlcolor=Black, linkcolor=blue,citecolor=blue]
 {hyperref}
\begin{document}

\title{
Combining Squeezing and Transition Sensitivity Resources for Quantum Metrology by Asymmetric Non-Linear Rabi model
}
\author{Zu-Jian Ying}
\email{yingzj@lzu.edu.cn}
\affiliation{School of Physical Science and Technology, Key Laboratory for Quantum Theory and Applications of MoE, and Lanzhou Center for Theoretical Physics, Lanzhou University, Lanzhou 730000, China}

\begin{abstract}
Squeezing and transition criticality are two main sensitivity resources for quantum metrology (QM), combination of them may yield an upgraded metrology protocol for higher upper bound of measurement precision (MP). We show that such a combination is feasible
in light-matter interactions by a realizable asymmetric non-linear quantum Rabi model (QRM). Indeed, the non-linear coupling possesses a squeezing resource for diverging MP while the non-monotonous degeneracy lifting by the asymmetries induces an additional tunable transition which further enhances the MP by several orders, as demonstrated by the quantum Fisher information. Moreover, the protocol is immune from the problem of diverging preparation time of probe state that may hinder 
the conventional linear QRM in application of QM. This work establishes a paradigmatic case of combining different sensitivity resources to manipulate QM and maximize MP.
\end{abstract}
\pacs{ }
\maketitle


{\it Introduction.}--The past two decades have witnessed both theoretical \cite%
{Braak2011,Solano2011,Boite2020,Liu2021AQT} and experimental progresses~\cite%
{ Diaz2019RevModPhy,Kockum2019NRP} in light-matter interactions\cite{Eckle-Book-Models,JC-Larson2021}, opening a
frontier field for explorations of exotic quantum states and developments of
quantum technologies. In particular, light-matter interactions manifest
finite-component quantum phase transitions (QPTs)~\cite{
Liu2021AQT,Ashhab2013,Ying2015,Hwang2015PRL,Ying2020-nonlinear-bias,Ying-2021-AQT,LiuM2017PRL,Hwang2016PRL,Irish2017, Ying-gapped-top,Ying-Stark-top,Ying-Spin-Winding,Ying-2018-arxiv,Ying-JC-winding,Ying-Topo-JC-nonHermitian,Ying-Topo-JC-nonHermitian-Fisher,Ying-gC-by-QFI-2024, Grimaudo2022q2QPT,Grimaudo2023-Entropy,Zhu2024PRL}
which have great application potential for critical quantum metrology (QM) with
high measurement
precision (MP)~\cite{Garbe2020,Garbe2021-Metrology,Ilias2022-Metrology,Chu2021-Metrology,Ying2022-Metrology,Montenegro2021-Metrology,Hotter2024-Metrology}.

Indeed, amidst the emerging phenomenology~\cite{LiPengBo-Magnon-PRL-2024,Qin-ExpLightMatter-2018,Li2020conical,Batchelor2015,
ChenQH2012,PengJ2021PRL,PengJie2019,PRX-Xie-Anistropy,Irish-class-quan-corresp,Irish2014,JC-Larson2021,WangZJ2020PRL,Lu-2018-1,
Ashhab2013,Ying2015,Hwang2015PRL,Ying2020-nonlinear-bias,Ying-2021-AQT,LiuM2017PRL,Hwang2016PRL,Irish2017, Ying-gapped-top,Ying-Stark-top,Ying-Spin-Winding,Ying-2018-arxiv,Ying-JC-winding,Ying-Topo-JC-nonHermitian,Ying-Topo-JC-nonHermitian-Fisher,Ying-gC-by-QFI-2024, Grimaudo2022q2QPT,Grimaudo2023-Entropy,Zhu2024PRL,
Cong2022Peter,Eckle-2017JPA,Ma2020Nonlinear,Yan2023-AQT,Lyu24-Multicritical,
Zheng2017,ZhengHang2017,Padilla2022,Yimin2018,Wu24-RabiTransition-Exp,
Gao2022Rabi-dimer,Gao2022Rabi-aniso,
Simone2018,FelicettiPRL2020,Alushi2023PRX,
LiuGang2023,YPWang2023QuanContrMag,AiQ2023,KuangLM2024AQT,
DiBello2024,Zhu-PRL-2020,Felicetti2015-TwoPhotonProcess,e-collpase-Garbe-2017,Rico2020,e-collpase-Duan-2016,CongLei2019,Felicetti2018-mixed-TPP-SPP,
Boite2016-Photon-Blockade,Ridolfo2012-Photon-Blockade,Casanova2018npj,Braak2019Symmetry} finite-component QPTs exhibit exotic properties of
criticality and universality~\cite{Ashhab2013,Ying2015,Hwang2015PRL,LiuM2017PRL,Hwang2016PRL,Irish2017,Ying-2021-AQT,Ying-Stark-top}, multi-criticality~\cite{Ying2020-nonlinear-bias,Ying-2021-AQT,Ying-gapped-top,Ying-Stark-top,Ying-2018-arxiv,
Ying-JC-winding,Zhu2024PRL,Lyu24-Multicritical,Wu24-RabiTransition-Exp},
compromise of universality and diversity~\cite{Ying-2021-AQT,Ying-Stark-top},
topological phase transitions with~\cite{Ying-2021-AQT,Ying-gapped-top,Ying-Stark-top} and without~\cite{Ying-gapped-top,Ying-Stark-top,Ying-Spin-Winding,Ying-JC-winding} gap closing, spin knot states~\cite{Ying-Spin-Winding},
coexistence of 
Landau-class and topological-class transitions~\cite{Ying-2021-AQT,Ying-Stark-top,Ying-JC-winding,Ying-Topo-JC-nonHermitian-Fisher},
robust topological feature against non-Hermiticity~\cite{Ying-Topo-JC-nonHermitian} and
universal criticality of exceptional points~\cite{Ying-Topo-JC-nonHermitian-Fisher}.
In reality, in the contemporary era of ultra-strong coupling~\cite{Ciuti2005EarlyUSC,Aji2009EarlyUSC,Diaz2019RevModPhy,Kockum2019NRP,Wallraff2004,Gunter2009, Niemczyk2010,Peropadre2010,FornDiaz2017,Forn-Diaz2010,Scalari2012,Xiang2013,Yoshihara2017NatPhys,Kockum2017,Ulstrong-JC-2,Ulstrong-JC-3-Adam-2019,PRX-Xie-Anistropy}
and deep-strong coupling~\cite{Yoshihara2017NatPhys,Bayer2017DeepStrong,Ulstrong-JC-1,DeepStrong-JC-Huang-2024} finite-component QPTs are
experimentally accessible~\cite{Yimin2018,Wu24-RabiTransition-Exp,Chen-2021-NC}.

A most promising application of the finite-component QPTs lies in
QM~\cite{Garbe2020,Garbe2021-Metrology,Ilias2022-Metrology,Chu2021-Metrology,Ying2022-Metrology,Montenegro2021-Metrology,Hotter2024-Metrology}, with the advantage of
high controllability and free of difficulty of reaching the equilibrium in
thermodynamical systems. 
Without doubt the main goal in QM is to raise MP as much as possible. Generally speaking there are various
resources for MP~\cite{Degen2017-QuantSensing}, including squeezing~\cite{Maccone2020Squeezing,Lawrie2019Squeezing,Gietka2023PRL-Squeezing,Gietka2023PRL2-Squeezing}, entanglement~\cite{Pezze2018entangelment} and
transition criticality~\cite{Garbe2020,Garbe2021-Metrology,Ilias2022-Metrology,Chu2021-Metrology,Ying2022-Metrology,Montenegro2021-Metrology,Hotter2024-Metrology}.
These resources individually can achieves high MP.
One may wonder about the possibility of combining different
resources.  Another issue concerned in QM is the reduction of preparation time of the probe
state (PTPS)~\cite{Garbe2020,Ying2022-Metrology,Gietka2022-ProbeTime}. In fact the PTPS in the linear quantum Rabi model (QRM) is diverging in the low-frequency
limit where the QPT occurs~\cite{Garbe2020,Ying2022-Metrology}, which may hinder the application for QM in
practice. Protocols with higher MP but finite PTPS are more favorable and desirable.

In the present work, rather than in the conventional low-frequency limit we reveal a tunable QPT in the low-tunneling limit. We show that the resource combination for QM can be realized in light-matter interactions by an asymmetric non-linear quantum Rabi model. As demonstrated by the quantum Fisher information (QFI), the non-linear coupling possesses a squeezing resource for diverging MP while the asymmetry competition introduces the additional QPT which raises the MP more by several orders. Favorably, the protocol has finite PTPS due to finite frequency and gap.

\begin{figure*}[t]
\includegraphics[width=1.88\columnwidth]{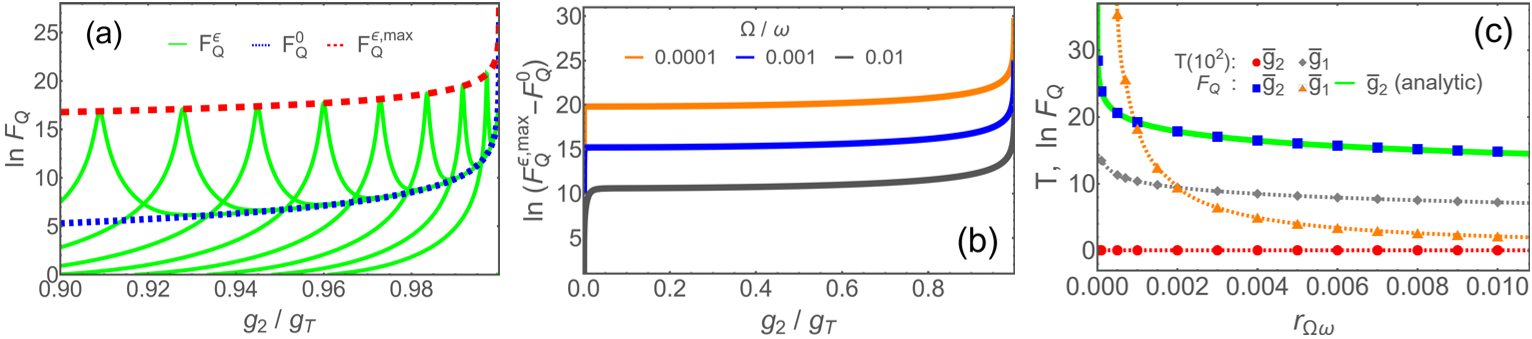}
\caption{Upgraded quantum metrology by combined sensitivity resources:
(a) QFI in natural logarithm at $\epsilon =0$ ($F_Q^0$, dotted), $\epsilon =0.27\sim0.34\omega$ by spacing $0.01\omega$  ($F_Q^\epsilon$, solid), and peak values at finite $\epsilon$  ($F_Q^{\epsilon,{\rm max}}$, dashed), for $\Omega/\omega=0.001$.
(b) Additional QFI gained by finite $\epsilon$ relative to zero-$\epsilon$ case at different $\Omega/\omega$ ratios.
(c) T in unit of $10^2$ for non-linear coupling (dots) compared with linear coupling (triangles)
and the maximum $F_Q$ for parameters $\overline{g}_2$ (squares) and $\overline{g}_1$\cite{Ying-gC-by-QFI-2024} (diamonds). The solid line is analytic summation of \eqref{F-transit-analytic} and \eqref{F-squeez-analytic} and $r _{\Omega\omega}  =\Omega/\omega$ ($\omega/\Omega$) for non-linear (linear) case in (c). $max(\Omega,\omega)=1$ is set as the unit.
}
\label{fig-QFI-combined}
\end{figure*}
\begin{figure*}[t]
\includegraphics[width=1.9\columnwidth]{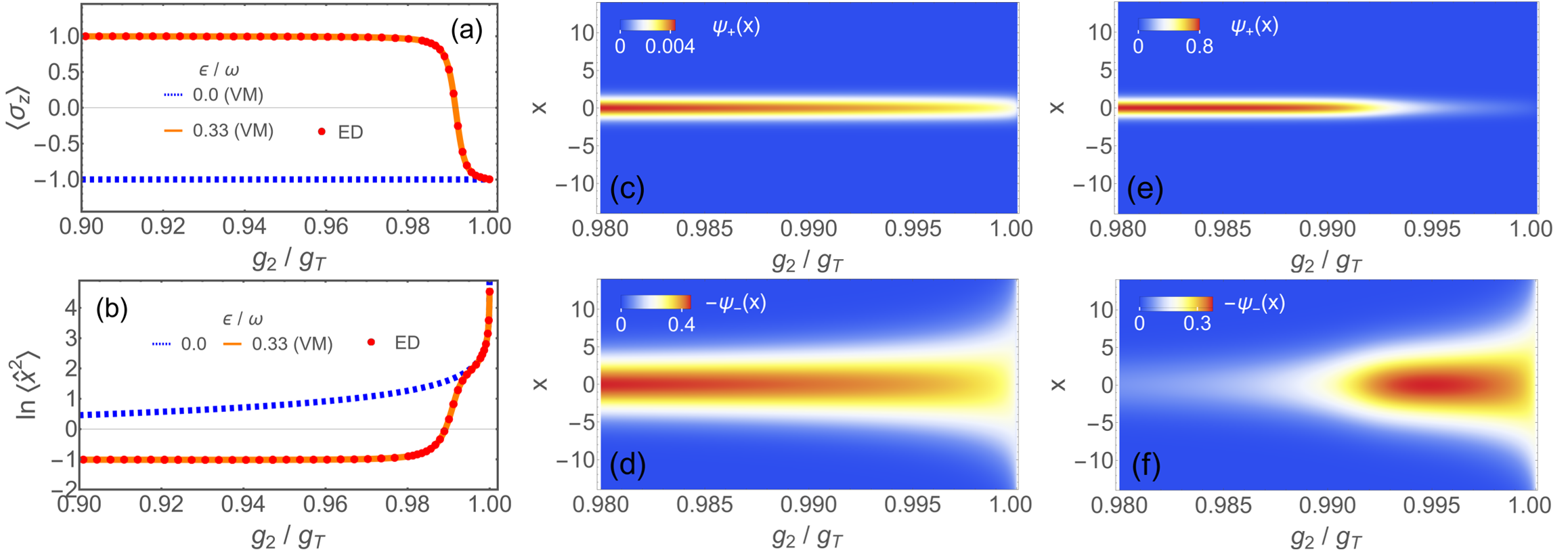}
\caption{Squeezing and transition sensitivity resources:
(a,b) $\langle \hat{\sigma} _x\rangle$ and $\langle \hat{x} ^2\rangle$ at $\epsilon =0$ (dotted) and $\epsilon =0.33\omega$ (solid).
(b-f) wave function $\psi _{\pm}$ at $\epsilon =0$ (b,c) and $\epsilon =0.33$ (e,f). Here $\Omega=0.01\omega$.  As illustrated in (a,b), variational method (VM, lines) and exact diagonalization (ED, dots)~\cite{Ying2020-nonlinear-bias} basically yield the same result.
}
\label{fig-wave-F}
\end{figure*}

\begin{figure*}[t]
\includegraphics[width=2\columnwidth]{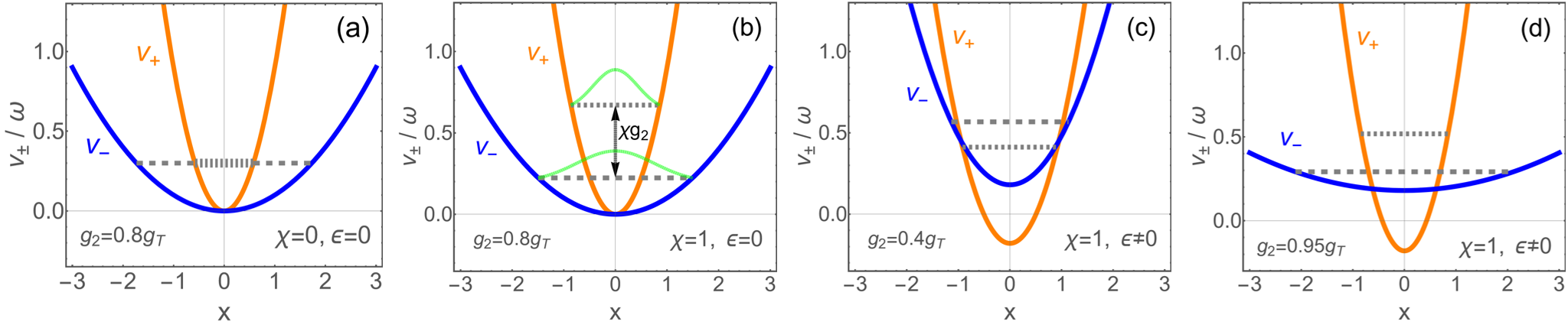}
\caption{Non-monotonous degeneracy lifting and origin of coupling-driven transition: $v_{\pm}$, $\varepsilon ^0 _{+}$ (dotted) and $\varepsilon ^0 _{-}$ (dashed) at
(a) $\chi=0$ and $\epsilon=0$,
(b) $\chi=1$ and $\epsilon=0$,
(c) $\chi=1$ and $\epsilon \neq 0$ with a small $g_2$,
(d) $\chi=1$ and $\epsilon \neq 0$ with a large $g_2$. The level splitting by $\chi g_2$ (arrows) and wave packets [green (light gray)] are denoted in (b).
}
\label{fig-v-potential}
\end{figure*}
\begin{figure*}[t]
\includegraphics[width=2\columnwidth]{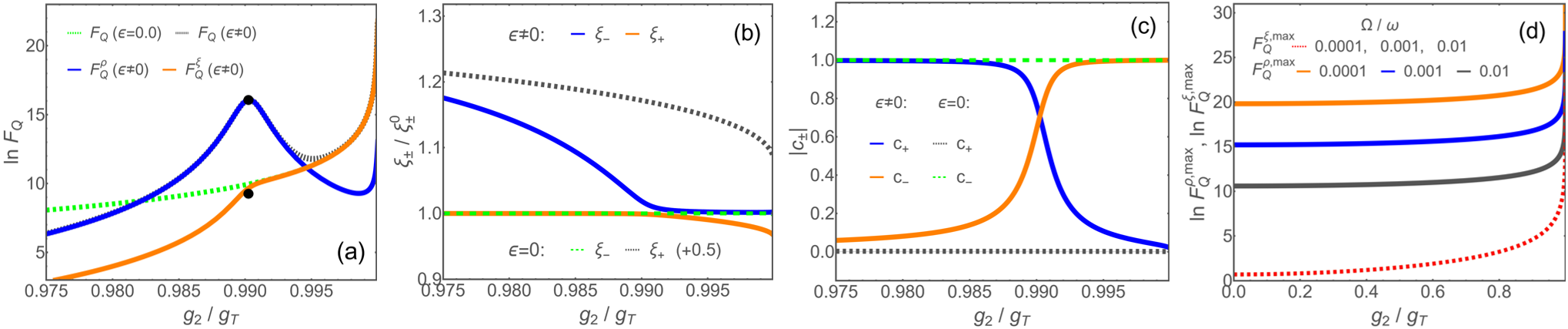}
\caption{Tracking the contributions of squeezing and transition.
(a) $F^\rho_Q$ [blue (dark gray) solid] and $F^\xi_Q$ [orange (light gray) solid] in total $F_Q$ at $\epsilon =0.328\omega$ compared with $\epsilon =0$ (dotted). The black dots denote analytic outcome of \eqref{F-transit-analytic} and \eqref{F-squeez-analytic} at peak position.
(b,c) variational $\xi _{\pm}$ (b) and $c_{\pm}$ (c) at $\epsilon =0$ (broken lines) and $\epsilon =0.328\omega$ (solid). $\xi _{+}/\xi^0 _{-}$ is shifted up by $0.5$ in (b) and
$\Omega=0.005\omega$ in (a-c). (d) $F^\rho_Q$ (solid) and $F^\xi_Q$ (dotted) at various ratios of $\Omega/\omega$.
}
\label{fig-Fq-apart}
\end{figure*}

{\it Model and asymmetries.}--%
We consider a non-linear QRM
\begin{equation}
H=\omega a^{\dagger }a+\frac{\Omega }{2}\hat{\sigma}_{x}+g_{2}\hat{\sigma}%
_{z}(a^{\dagger }+a)^{2}-\epsilon \hat{\sigma}_{z}  \label{H-T-bias}
\end{equation}%
which describes a quadratic coupling between a bosonic mode with frequency $%
\omega $, created (annihilated) by $a^{\dagger }$ ($a)$, and a qubit
represented by the Pauli matrices $\hat{\sigma}_{x,y,z}$. As a comparison
the linear QRM~\cite{rabi1936,Rabi-Braak,Eckle-Book-Models} has a coupling in form of $g_{1}\hat{\sigma}_{z}(a^{\dagger
}+a)$. Here the quadratic coupling $\hat{\sigma}_{z}(a^{\dagger }+a)^{2}$
can be rewritten as\cite{Ying-2018-arxiv,Ying2020-nonlinear-bias}
\begin{equation}
\hat{\sigma}_{z}[(a^{\dagger })^{2}+a^{2}]+\chi \hat{\sigma}_{z}(2a^{\dagger
}a+1)
\end{equation}%
with $\chi =1$, while the $\chi =0$ case is called the two-photon QRM~\cite{Felicetti2018-mixed-TPP-SPP,Felicetti2015-TwoPhotonProcess,e-collpase-Garbe-2017,Rico2020,e-collpase-Duan-2016,CongLei2019} and $\sigma _{z}a^{\dagger }a$ is
a Stark-like term~\cite{Eckle-2017JPA}. Although the $\chi =0$ case is more symmetric with symmetry
$P_{2}=\sigma _{x}e^{i\pi a^{\dagger }a/2}$, the full quadratic form
with $\chi =1$ in (\ref{H-T-bias}) is more original in circuit systems~\cite{Felicetti2018-mixed-TPP-SPP}.
While the bias $\epsilon $ is usually responsible for
asymmetry~\cite{Braak2011,HiddenSymLi2021,HiddenSymBustos2021,HiddenSymMangazeev2021,Ying-2018-arxiv,Ying2020-nonlinear-bias},
the $\chi $ term introduces a new asymmetry origin to break the $P_2$ symmetry.
Another symmetry $P_{x}=e^{i\pi a^{\dagger }a}$~\cite{Ying-2021-AQT,Ying-JC-winding} takes the place in $H$, however 
it lifts the spin degeneracy imposed by $P_2$.
Such a symmetry variation releases a non-monotonous degeneracy lifting and brings about a first-order-like transition which provides the
transition sensitivity resource in addition to the squeezing resource for
the combined QM, as we shall address in this work.

For the convenience of further analysis, by transformation
$a^{\dagger }=(\hat{x}-i\hat{p})/\sqrt{2},$ $a=(\hat{x}+i\hat{p})/\sqrt{2}$
with position $x$ and momentum $\hat{p}=-i\frac{\partial }{\partial x}$, we
can rewrite $H$ as~\cite{Irish2014,Ying2015,Ying-2018-arxiv,Ying2020-nonlinear-bias}
$H_{x}=\sum_{\sigma _{z}=\pm }h_{\sigma _{z}}\left\vert \sigma
_{z}\right\rangle \left\langle \sigma _{z}\right\vert +\frac{\Omega }{2}
\sum_{\sigma _{z}=\pm }\left\vert \sigma _{z}\right\rangle \left\langle
\overline{\sigma }_{z}\right\vert $ where $\sigma _{z}=-\overline{\sigma }
_{z}=\pm $ labels the spin in $z$ direction. Here $h_{\pm }=\frac{%
\omega }{2m_{\pm }}\hat{p}^{2}+v_{\pm }\left( x\right) -\frac{1}{2}\omega $
is the effective singe-particle Hamiltonian, in the spin-dependent harmonic
potential
\begin{equation}
v_{\pm }(x)=\frac{\omega }{2}m_{\pm }\varpi _{\pm }^{2}x^{2}\mp \epsilon ,
\label{v-potentials}
\end{equation}%
with effective mass $m_{\pm }=\left[ 1\mp \widetilde{\chi }\overline{g}_{2}%
\right] ^{-1}$, renormalized frequency $\varpi _{\pm }=\sqrt{\left( 1\pm
\overline{g}_{2}\right) \left( 1\mp \widetilde{\chi }\overline{g}_{2}\right)
}$ and single-particle energy $\varepsilon _{\pm }=\varepsilon _{\pm
}^{0}\mp \epsilon -\frac{1}{2}\omega $, where%
\begin{equation}
\varepsilon _{\pm }^{0}=\omega (n+\frac{1}{2})\sqrt{1\pm \left( 1-\widetilde{%
\chi }\right) \overline{g}_{2}-\widetilde{\chi }\overline{g}_{2}^{2}}.
\label{e0-up-down}
\end{equation}%
Here
$\widetilde{\chi }=\left( 1-\chi \right) /\left( 1+\chi \right) $,
$\overline{g}_{2}=g_{2}/g_{\mathrm{T}} \leqslant 1$ before the spectral collapse point $g_{\mathrm{T}}=\omega /[2\left(
1+\chi \right) ]$~\cite{Ying-2018-arxiv,Ying2020-nonlinear-bias,Felicetti2018-mixed-TPP-SPP,Felicetti2015-TwoPhotonProcess,e-collpase-Garbe-2017,Rico2020,e-collpase-Duan-2016,CongLei2019}. In such a formalism the $\Omega $ term plays the role of
spin flipping in the spin space or tunneling in the position space~\cite{Irish2014,Ying2015,flux-qubit-Mooij-1999}.
Hereafter we consider the ground state which has $n=0$.

{\it Diverging QFI in $\protect\epsilon =0$ case.}--In QM the MP of experimental estimation on a parameter
$\lambda $ is bounded by $F_{Q}^{1/2}$\cite{Cramer-Rao-bound}, where $F_{Q}$
is the QFI defined as \cite%
{Cramer-Rao-bound,Taddei2013FisherInfo,RamsPRX2018}
\begin{equation}
F_{Q}=4\left[ \langle \psi ^{\prime }\left( \lambda \right) |\psi ^{\prime
}\left( \lambda \right) \rangle -\left\vert \langle \psi ^{\prime }\left(
\lambda \right) |\psi \left( \lambda \right) \rangle \right\vert ^{2}\right]
\label{Fq}
\end{equation}%
for a pure states $|\psi (\lambda )\rangle $. Here $^{\prime }$ denotes the derivative with respect to the parameter $\lambda $. A
higher QFI would mean a higher MP for QM. In the absence of
the bias, the QFI for the parameter $\lambda =g_{2}$ manifests a
diverging behavior when $g_{2}$ approaches $g_{%
\mathrm{T}}$, as demonstrated by the dotted line in Fig.\ref%
{fig-QFI-combined}(a). Such a diverging QFI comes from the squeezing effect
driven by the frequency renormalization $\varpi _{\pm }$. Indeed, as shown
in\ Fig.\ref{fig-wave-F}(c) and \ref{fig-wave-F}(d) the wave function $\psi
_{\pm }(x)$ is narrowing (widening) in the spin-up (spin-down) component,
corresponding to an amplitude (phase) squeezing~\cite{Ref-Squeezing}.
In particular, such a
squeezing effect is divergently strong in $\psi _{-}(x)$ as $\varpi _{-}$
is vanishing in approaching $g_{\mathrm{T}}$. Note that the single-particle energy in the ground state, $\varepsilon
_{\pm }=\frac{1}{2}\varpi _{\pm }\omega$, is lower in the spin-down
component. Thus $\psi _{-}(x)$ has a dominate weight in a low $\Omega
/\omega $ ratio, as indicated by the
spin expectation value, $\langle \hat{\sigma}_{z}\rangle \approx -1$,
denoted by the dotted line in Fig.\ref{fig-wave-F}(a). The accelerated
squeezing, especially in $\psi _{-}(x)$ in the vicinity of $g_{\mathrm{T}}$, provides
sensitivity resource for the diverging QFI.

{\it Upgraded QFI in $\protect\epsilon \neq 0$ case.}--The QFI can be much enlarged in the presence of the bias. Indeed, as
illustrated by Fig.\ref{fig-QFI-combined}(a), high $F_{Q}$ peaks (green
solid lines) emerge when finite values of bias are introduced. Note that
such peak values are higher by several orders than the $F_{Q}^{0}$ values in
the $\epsilon =0$ case (blue dotted line), thus providing a continuously
reachable upper bound of QFI values $F_{Q}^{\epsilon ,\max }$ (red dashed
line). By tuning to a lower $\Omega /\omega $ ratio, in finite $\epsilon $
cases one can gain a larger extra value of the QFI over the already diverging
$F_{Q}^{0}$, as shown by Fig.\ref{fig-QFI-combined}(b).

{\it Combined squeezing and transition sensitivity resources.}--Actually the QFI is also equivalent to the susceptibility of the fidelity
whose peak signals a QPT~\cite{Zhou-FidelityQPT-2008,Zanardi-FidelityQPT-2006,Gu-FidelityQPT-2010,You-FidelityQPT-2007,You-FidelityQPT-2015},
as applied to extract the accurate frequency dependence of the
second-order-like transition of the linear QRM~\cite{Ying-gC-by-QFI-2024}.
Here for the non-linear QRM the QFI peaks in $\epsilon \neq 0$ case also arise from a bias-tuned first-order-like
transition, as illustrated by the fast change in $\langle \hat{\sigma} _{z}\rangle $ (orange solid line) around
$g_{2}=0.99g_{\mathrm{T}}$ in Fig.\ref{fig-wave-F}(a). One can see that the transition indeed
adds a boost of property variation (orange solid line) to the original fast
changing from the squeezing effect (dotted line) in $\langle \hat{x}^{2}\rangle $
in Fig.\ref{fig-wave-F}(b). More essentially in the wave function,
the amplitude of $\psi _{+}(x)$ decreases quickly while that of $\psi _{-}(x)
$ grows fast around the transition, as in Figs.\ref{fig-wave-F}(e) and \ref%
{fig-wave-F}(f), in addition to the wave-packet narrowing and broadening of $%
\psi _{+}(x)$ and $\psi _{-}(x)$. Such a quicker variation in the wave
function resulting from both the transition and the squeezing yields a
combined sensitivity resource for the enlargements of $F_{Q}$ and upgraded MP.

{\it Transition point and optimal bias.}--The emerged transition in finite bias can be seen clearly by the formalism
in $H_{x}$. Firstly we stress that such a coupling-driven transition does not exist in the
conventional two-photon QRM ($\chi =0$ case) as we have always degenerate
single-particle energy $\varepsilon _{+}^{0}=$ $\varepsilon _{-}^{0}$ in Fig.\ref{fig-v-potential}(a), due
to $\widetilde{\chi }=1$ in such situation, while an added bias only lifts
the degeneracy monotonously which cannot bring a transition in coupling variation. In contrast, in $\chi =1$ case, the degeneracy is already
lifted in the absence of bias by the coupling, with $\varepsilon _{+}^{0}>$ $%
\varepsilon _{-}^{0}$, as in Fig.\ref{fig-v-potential}(b). In
the presence of bias, the two energy levels can be reversed in a small $g_{2}
$, with $\varepsilon _{+}^{0}<$ $\varepsilon _{-}^{0}$, as in Fig.\ref%
{fig-v-potential}(c), while they can be reversed again in a large $g_{2}$,
recovering $\varepsilon _{+}^{0}>$ $\varepsilon _{-}^{0}$, as in Fig.\ref%
{fig-v-potential}(d). The level crossing $\varepsilon _{+}=\varepsilon _{-}$
gives the transition point
\begin{equation}
g_{2c}=4\sqrt{\left( \epsilon /\omega \right) ^{2}-4\left( \epsilon /\omega
\right) ^{4}}g_{\mathrm{T}}
\end{equation}%
at a given bias. Since the transition point has the $F_{Q}$ peak, we can
tune the bias, for measurements at a coupling $g_{2}$, to the optimal one
$\epsilon _{\max }=\frac{\omega}{4} ( \sqrt{1+\overline{g}_{2}}-\sqrt{
1-\overline{g}_{2}} )  $
at which the $F_{Q}$ value reaches the maximum.

{\it QFI-contribution tracking for two resources.}--To have a better view of the contributions of the two sensitivity resources
to the QFI, we can assume $\psi _{\pm }(x)=c_{\pm }\varphi _{\pm }$ where
$\varphi _{\pm }=\xi _{\pm }^{1/4}\exp [-\frac{1}{2}\xi _{\pm }x^{2}]/\pi ^{1/4}$
is the ground state of harmonic oscillator with frequency
renormalization factor $\xi _{\pm }$ as in the variational method of polaron
picture\cite{Ying2015}. Here $c_{+}=(e_{-}+\eta \sqrt{e_{-}^{2}+S_{\Omega
}^{2}})/N$ and $c_{-}=S_{\Omega }/N$, with $e_{\pm }=(\widetilde{\varepsilon
}_{+}\pm \widetilde{\varepsilon }_{-})/2$ and $N$ subject to normalization
condition $c_{+}^{2}+c_{-}^{2}=1$. The final $\xi _{\pm }$ is determined by
the minimization of the variational energy $E^{-}$, where the ground state
takes the $\eta =-$ branch from $E^{\eta }=e_{+}+\eta \sqrt{%
e_{-}^{2}+S_{\Omega }^{2}}$ and $\widetilde{\varepsilon }_{\pm }=\left( 1\pm
\overline{g}_{2}+\xi _{\pm }^{2}\right) \frac{\omega }{4\xi _{\pm }}\mp
\epsilon $, $S_{\Omega }=\frac{\Omega }{2}\langle \varphi _{+}|\varphi
_{-}\rangle =(\xi _{+}\xi _{-})^{1/4}\Omega /\sqrt{2(\xi _{+}+\xi _{-})}$.

Note $\langle \psi ^{\prime }\left( \lambda \right) |\psi \left( \lambda
\right) \rangle =0$ for a real wave function\cite{Ying-gC-by-QFI-2024}, the
QFI is simplified to be $F_{Q}=4\langle \psi ^{\prime }\left( \lambda
\right) |\psi ^{\prime }\left( \lambda \right) \rangle $ which finally only
contains two parts: $F_{Q}=F_{Q}^{\rho }+F_{Q}^{\xi }$ where
\begin{equation}
F_{Q}^{\rho }=4\sum_{\sigma =\pm }\left( \frac{dc_{\sigma }}{dg_{2}}\right)
^{2},\quad F_{Q}^{\xi }=4\sum_{\sigma =\pm }c_{\sigma }^{2}S_{\varphi
^{\prime }\varphi ^{\prime }}^{\sigma ,\xi }\left( \frac{d\xi _{\sigma }}{dg_{2}}\right) ^{2},
\end{equation}
and $S_{\varphi ^{\prime }\varphi ^{\prime }}^{\pm ,\xi }=\langle \frac{%
d\varphi _{\pm }}{d\xi _{\pm }}|\frac{d\varphi _{\pm }}{d\xi _{\pm }}\rangle
=(8\xi _{\pm }^{2})^{-1}$. Apparently $F_{q}^{\xi }$ is the contribution
from the squeezing effect, while $F_{q}^{\rho }$ is the contribution from
the wave-packet weight variation in the transition. We find
that the mixed term of the two resources, with mixed variation factor $\frac{%
dc_{\sigma }}{dg_{2}}\frac{d\xi _{\sigma }}{dg_{2}}$, vanishes here as$\
\langle \varphi _{\sigma }^{\prime }|\varphi _{\sigma }\rangle =\langle
\varphi _{\sigma }|\varphi _{\sigma }\rangle ^{\prime }=0$ due to the basis
normalization $\langle \varphi _{\sigma }|\varphi _{\sigma }\rangle =1$. The
variation of $c_{\pm }$ is illustrated in Fig.\ref{fig-Fq-apart}(c), showing
a quick change around the transition $g_{2c}\approx 0.99g_{\mathrm{T}}$ at
the finite bias in contrast to the locally flat behavior at zero bias. The
evolution of $\xi _{\pm }$ rescaled by $\xi _{\pm }^{0}=\varpi _{\pm }$ is
plotted in Fig.\ref{fig-Fq-apart}(b), indicating that the variation of $\xi
_{\pm }$ mainly follows the scale of $\varpi _{\pm }$ despite of a
critical-like behavior before and after $g_{2c}$ for $\xi _{-}/\xi _{-}^{0}$
[blue (dark-gray) solid) and $\xi _{+}/\xi _{+}^{0}$ [orange (light-gray)
solid] respectively. The contributions of $F_{Q}^{\rho }$ [blue (dark-gray)
solid] and $F_{Q}^{\xi }$ [orange (light-gray) solid] at the finite bias are
tracked in Fig.\ref{fig-Fq-apart}(a), we see that $F_{Q}^{\xi }$ maintains basically
the same amount of QFI as in $\epsilon =0$ case (green dotted) after $g_{2c}$%
. There is some discount in $g<g_{2c}$ regime due to the smaller weight in $%
c_{-}$, which however doesnot affect the upper bound of QFI that is what we
need. On the other hand, from Fig.\ref{fig-Fq-apart}(a) we see that $%
F_{Q}^{\rho }$ indeed is responsible for the dramatic increase of the QFI.
Since the transition occurs around the degenerate point of the
single-particle energy, we can readily obtain in the leading order the two
contributions at $\epsilon _{\max }$ explicitly as
\begin{eqnarray}
F_{Q}^{\rho ,\max } &=&\frac{\left[ \overline{g}_{2}^{4}+4\left(
1+w_{2}\right) -\overline{g}_{2}^{2}\left( 5+3w_{2}\right) \right]
^{2}\omega ^{2}}{2w_{2}^{9/2}\left( w_{+}+w_{-}\right) ^{5}\left(
1+w_{2}\right) ^{2}\Omega ^{2}\ g_{\mathrm{T}}^{2}},
\label{F-transit-analytic} \\
F_{Q}^{\xi ,\max } &=&\left( 1+\overline{g}_{2}^{2}\right) /\left[ 8(1-%
\overline{g}_{2}^{2})^{2}g_{\mathrm{T}}^{2}\right] ,
\label{F-squeez-analytic}
\end{eqnarray}%
where $w_{2}=w_{+}w_{-}$and $w_{\pm }=\sqrt{1\pm \overline{g}_{2}}$. The
analytic results of $F_{Q}^{\rho ,\max }$ and $F_{Q}^{\xi ,\max }$ agree
well with the numerics, as denoted by the black dots in Fig.\ref%
{fig-Fq-apart}(a) [also squares in Fig.\ref{fig-QFI-combined}(c)]. We see that one gets a diverging $F_{Q}^{\xi ,\max }$ in
proximity of $g_{2}=g_{\mathrm{T}}$ universally for small ratios of $\Omega
/\omega $, while diverging $F_{Q}^{\rho ,\max }$ is available overall $g_{2}$
regime with respect to small ratios of $\Omega /\omega $, as shown in Fig.%
\ref{fig-Fq-apart}(d). Around $g_{\mathrm{T}}$ the tunable $F_{Q}^{\rho
,\max }$ is larger than $F_{Q}^{\xi ,\max }$ as long as $\Omega /\omega <(%
\frac{1-\overline{g}_{2}}{2})^{3/8}$, e.g., $\Omega /\omega =0.01$ ($0.001)$
guarantees a larger $F_{Q}^{\rho ,\max }$ for $\overline{g}_{2}<0.999991$ ($%
0.99999998)$ which actually covers the entire practical regime.

{\it Gap and PTPS.}--Another issue concerned in QM is the PTPS, which is inversely
proportional to the instantaneous gap as estimated by $T=\int_{0}^{\overline{%
g}_{2c}}\Delta \left( \overline{g}_{2}\right) ^{-1}d\overline{g}_{2}$. The
conventional linear QRM has a second-order QPT in
the opposite limit $\omega /\Omega \rightarrow 0$~\cite{
Ashhab2013,Ying2015,Hwang2015PRL,LiuM2017PRL,Irish2017} applied for QM~\cite{Garbe2020},
the PTPS is however diverging~\cite{Garbe2020,Ying2022-Metrology} as the exponentially closing gap is overall in
order of $\omega $\cite{Ying-2021-AQT,Ying2022-Metrology}. In contrast, for the non-linear model here, the gap
$\Delta \left( \overline{g}_{2}\right) =E^{+}-E^{-}=2\sqrt{
e_{-}^{2}+S_{\Omega }^{2}}$ has an order of $\left( w_{+}-w_{-}\right)
\omega /2-2\epsilon $ apart from an unclosed minimum gap around $g_{2c}$ and does not overall vanish due to finite $\omega$ and $\epsilon$.
Such a gap situation guarantees an always finite PTPS. Indeed, as compared in
Fig.\ref{fig-QFI-combined}(c) with $\overline{g}_{2c}=0.999$, the PTPS in non-linear case (dots) is only a
single digit while it is diverging in linear case (triangles). We have not
mentioned that the QFI is also higher by orders, as shown by the squares
(non-linear) and diamonds (linear) in Fig.\ref{fig-QFI-combined}(c). Thus
our non-linear  protocol is more favorable in practice than the linear one
in applications for QM.

{\it Conclusion.}--With the asymmetric non-linear QRM realizable in circuit systems
we have explored an upgraded
protocol of QM by a combination of squeezing and transition
sensitivity resources. The tunable asymmetry-induced transition adds a dramatic
increase of QFI over the already-diverging QFI from the squeezing resource,
indicating an improvement of MP by orders. In addition, the protocol is
immune from the problem of diverging
PTPS which exists in the conventional linear QRM in application of QM.
We have clarified the mechanism for the improvement and obtained
analytic results of the maximum QFI and the optimal bias. Our work
establishes a paradigmatic case of combining different sensitivity resources
to upgrade QM in light-matter interactions. A broader
combination to include entanglement resource by adding more qubits and bosonic modes
might be even more advantageous, which can be a further work.

{\bf Acknowledgment.}--This work was supported by the National Natural Science Foundation of China
(Grants No. 12474358, No. 11974151, and No. 12247101).


\end{document}